\documentclass[aps,reply,twocolumn,showpacs,showkeys,nofootinbib,superscriptaddress]{revtex4-1}

\usepackage[utf8]{inputenc}
\usepackage{graphicx}
\usepackage{amsfonts}
\usepackage{amssymb}
\usepackage{amsmath}
\usepackage[mathscr]{eucal}
\usepackage{setspace}
\usepackage{booktabs}
\usepackage[shellescape,latex]{gmp}
\usempxpackage{amssymb}

\def\sqr#1#2{{\vcenter{\hrule height.#2pt
   \hbox{\vrule width.#2pt height#1pt \kern#1pt
      \vrule width.#2pt}
   \hrule height.#2pt}}}

\def\bsqr#1#2{{\vrule width #1pt height#2pt}}
\def\bsquare{{\mathchoice\bsqr66\bsqr66\bsqr33\bsqr33}}
\def\badbreak{\penalty1000}

\def\union{\cup}                             
\newcommand{\cP}{{\cal P}}                    
\newcommand{\cC}{{\cal C}}                    
\def\fir{{\scriptscriptstyle{\text{\rm IR}}}}             
\def\fuv{{\scriptscriptstyle{\text{\rm UV}}}}             
\def\smax{{\scriptstyle{\text{\rm max}}}}          
\def\smin{{\scriptstyle{\text{\rm min}}}}            
\def\lm0{{\lambda_0}}                                     
\def\nrN{N}                                               
\def\efN{\mathscr{N}}                                     
\def\efNm{\efN_\star}                                     
\def\v{b}                                                            

\makeatletter
\newcommand*{\smallrel}[2][.8]{%
  \mathrel{\mathpalette{\smallrel@{#1}}{#2}}%
}
\newcommand*{\smallrel@}[3]{%
  \sbox0{$#2\vcenter{}$}%
  \dimen@=\ht0 %
  \raise\dimen@\hbox{%
    \scalebox{#1}{%
      \raise-\dimen@\hbox{$#2#3\m@th$}%
    }%
  }%
}
\makeatother

\usepackage{amsmath} 
\usepackage{dsfont}  
\usepackage{bm}      

\def\beq{\begin{equation}}
\def\eeq{\end{equation}}
\def\beqs#1\eeqs{\beq\begin{split} #1 \end{split}\eeq}

\long\def\comment#1{}

\def\be{\begin{equation}}
\def\ee{\end{equation}}

\def\bc{\begin{center}}
\def\ec{\end{center}}

\usepackage{microtype}
\usepackage[dvipsnames]{xcolor}
\usepackage[colorlinks=true,backref=false, linktocpage=true,
citecolor=PineGreen,urlcolor=PineGreen,linkcolor=PineGreen,pdfpagemode=UseOutlines]{hyperref}

\hypersetup{%
  bookmarksnumbered=true,
  pdftitle = {},
  pdfsubject = {},
  pdfauthor = {},
  pdfkeywords = {}
}

\begin{document}

\title{Response to Comment arXiv:2210.10539v2}

\author{Ivan Horv\'{a}th}
\email{ihorv2@g.uky.edu}
\affiliation{Nuclear Physics Institute CAS, 25068 \v{R}e\v{z} (Prague), Czech Republic}
\affiliation{University of Kentucky, Lexington, KY 40506, USA}

\author{Peter Marko\v{s}}
\email{peter.markos@fmph.uniba.sk}
\affiliation{Dept. of Experimental Physics, Faculty of Mathematics, 
Physics and Informatics, Comenius University in Bratislava, Mlynsk\'a Dolina 2, 
842 28 Bratislava, Slovakia}

\date{Feb 28, 2023}

\maketitle 

\noindent
In recent work~\cite{Horvath:2021zjk}, we calculated the infrared (IR) effective
counting dimension 
$d_\fir$~\cite{Horvath:2018aap, Alexandru:2021pap, Horvath:2022ewv}
for critical states of $3D$ Anderson models in universality classes O, U, S 
and AIII. The results entailed two new messages. 
{\em (m1)} Space effectively occupied by critical electron is of dimension 
$d_\fir \!\approx\! 8/3$. Demonstrated properties of effective counting imply 
that the meaning and relevance of this is fully analogous to e.g. Minkowski 
dimension of ternary Cantor set being $d_\fuv \!=\! \log_3(2)$. Indeed, both 
statements express the scaling of properly defined physical volume; both 
dimensions are measure-based. 
{\em (m2)} The values of $d_\fir$ in studied classes coincide to better than two 
parts per mill with comparable errors. We dubbed this finding superuniversality 
of $d_\fir$ since other critical indices tend to differ to a notably larger degree. 
Exact superuniversality offers itself as a possibility, but obviously cannot 
be demonstrated in a finite numerical computation, only ascertained 
with better bounds or disproved.

In Ref. \cite{Burmistrov:2022} Burmistrov claims that he ``proved" inexact nature 
of {\em (m2)} by invoking the multifractal (MF) formalism. He derived 
MF representation for average effective count $\langle \efNm \rangle$, 
involving $f(3)$ and $\alpha_0$ ($f(\alpha_0) \!=\!3$),~namely
\begin{equation}
     \langle \efNm \rangle \simeq 4 c_0 \frac{L^{f(3)}}{\sqrt{\ln(L)}}  \quad,\quad
     c_0=\sqrt{| f''(\alpha_0) | / (2\pi)}
     \label{eq:805}
\end{equation}
The asymptotically equal sign ($\,\simeq\,$) suggests that \eqref{eq:805} 
conveys an exact $L \!\to\! \infty$ leading term of $\langle \efNm \rangle$.
However, in the ensuing discussion, $4c_0$ appears to be treated as  
approximation to proportionality constant, which is what we will assume. 
Burmistrov then concludes that~\eqref{eq:805} 
{\em ``proves the absence of ``superuniversality" of $\langle \efNm \rangle$"} 
due to numerically known class-dependent values of $f(3)$ and $c_0$.
We note in passing that superuniversality of $\langle \efNm \rangle$,
exact or not, was in fact not claimed or invoked in~\cite{Horvath:2021zjk}.

To put formula~\eqref{eq:805} in context, recall that MF formalism was 
created to describe UV measure singularities arising e.g. in strange attractors 
\cite{falconer2014fractal, Halsey_Kadanoff_multif:1986}. The method 
identifies sets $A_\alpha$ of local singularities with H\"older strength 
$\alpha \!>\!0$, treating their Hausdorff dimensions 
$f_H(\alpha) \!=\! \text{dim}_H(A_\alpha)$ as characteristics of interest. 
Note that $f_H(\alpha)$ is a proper measure-based dimension 
of spatial set $A_\alpha$. 

Common variation is the moment method which avoids 
computing $\alpha(x)$ at each point for the price of coarsening the singularity 
information. Evaluation of the associated $f_m(\alpha)$ proceeds by computing 
continuum of generalized dimensions which are not measure-based 
and do not represent dimensions of space in themselves. Nevertheless, 
$f_H(\alpha)$ and $f_m(\alpha)$ coincide under certain analytic assumptions 
by virtue of associated transformations. However, the moment method does not
identify the singularity set $A_\alpha$. What it does provide is a trick to evaluate 
$f_H(\alpha)$ valid in certain circumstances. Without explicit check that
$f_H(\alpha) \!=\! f_m(\alpha)$ or explicit verification that needed analytic 
properties are met, one has no choice but to treat $f_m(\alpha)$ as 
approximation to $f_H(\alpha)$.
 
In Anderson criticality, the situation is somewhat different. This is 
an IR problem which means that the notion of local singularity is absent and 
the very concept of $A_\alpha$ becomes fuzzy. The moment method is then 
normally taken to {\em define} MF spectrum $f(\alpha) \!\equiv\! f_m(\alpha)$ 
upon formally replacing the UV cutoff $a$ with $1/L$. However, one is left 
with little regarding the information on populations (subsets of space) whose 
dimensions is $f_m(\alpha)$ meant to represent. Unlike in UV problem of 
a strange attractor, there is no readily available analogue of $A_\alpha$ 
and $f_H(\alpha)$ to independently check upon.

Formula \eqref{eq:805} may prove valuable in this regard since it represents 
MF {\em prediction} ($f \!\equiv\! f_m$) for $\langle \efNm \rangle$ which 
has independent definition and well-defined spatial meaning. Degree of 
consistency between true $\langle \efNm \rangle$ and \eqref{eq:805} can 
thus test assumptions associated with the MF approach. We perform 
such comparison to 
$\langle \efNm \rangle$ data of Ref.~\cite{Horvath:2021zjk}. To minimize 
finite-$L$ systematics, only the upper segment of all data ($64 \le L \le 128$) 
is used to fit for $c_0$ and $f(3)$. We find that instead
of improving, fit form \eqref{eq:805} worsens $\chi^2$/dof relative 
to pure power ($1.5 \rightarrow 2.9$). More importantly, resulting $f(3)$ 
is sharply inconsistent with the MF-computed value $f(3)\!=\! 2.733(3)$ 
\cite{Ujfalusi:2015a}, used in Burmistrov argument. Described results
refer to O class but situation in other classes is the same.

\begin{figure}[t]
   \includegraphics[width=0.45\textwidth]{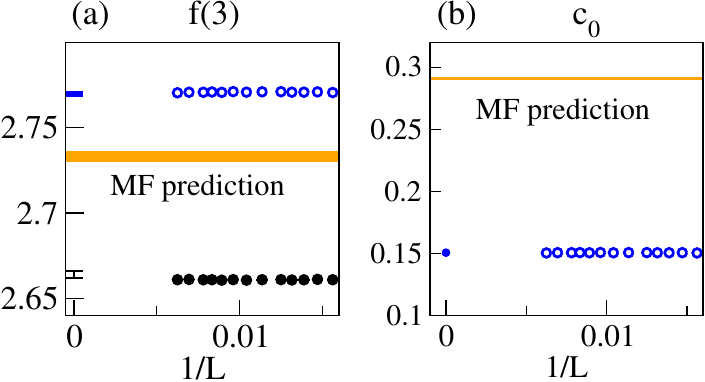}
   \vskip -0.05in   
   \caption{MF-computed indices $f(3)$ \cite{Ujfalusi:2015a}  and $c_0$ 
   compared to those obtained from Eq.~\eqref{eq:805} and  
   $\langle \efNm \rangle$ data of Ref.~\cite{Horvath:2021zjk}.}  
   \label{fig:Burm1}
   \vskip -0.16in
\end{figure}

The above is conveyed by Fig.~\ref{fig:Burm1}(a) where $1/L$ behavior 
of $f(3)$, expressed from  Eq.~\eqref{eq:805}, is shown together with 
fitted $1/L \!=\!0$ value (with error) and the associated MF prediction. 
Analogous plot for pure power is also shown (full circles). Fitted $d_\fir$
here is consistent with one obtained from the extrapolated pair 
method~\cite{Horvath:2021zjk}. Approach of data to fitted values in both
cases is of course by design, similarly to plots in Fig.~1 of 
Ref.~\cite{Burmistrov:2022}. 
Note also that the obtained $c_0$ differs roughly by factor of two 
from its approximate MF prediction as shown in Fig.~\ref{fig:Burm1}(b).

\noindent
The above discussion warrants the following points.

\smallskip

\noindent
{\em (i)} Given the scale of inconsistencies found in the MF prediction~\eqref{eq:805}, 
we conclude that arguments of Ref.~\cite{Burmistrov:2022} do not have tangible
bearing on our {\em (m1)} and {\em (m2)}, including the possibility of exact 
superuniversality in $d_\fir$. One should also keep in mind that superuniversality 
and $d_\fir \!\approx\! 8/3$ are two distinct conclusions.  

\vspace*{0.02in}
\noindent
{\em (ii)} Our findings suggest that derivation of \eqref{eq:805} may involve 
assumptions on $\alpha$-populations that are not justified. For example, setting 
$p(x) \!\equiv\! L^{-\alpha(x)}$ ($p(x)=\,$probability at $x$), which forces certain 
structure on $\alpha$-populations without knowing what they actually are, may be 
too cavalier. Showing that this is harmless in $L \!\to\! \infty$ limit is non-trivial even 
for moderately complex populations.
Similar goes for setting $\cP(\alpha) \!\equiv\! \cC L^{f(\alpha)}$ where $\cP(\alpha)$ 
is the $\alpha$-population count.
Assumptions on analyticity of $f(\alpha)$, $f_m(\alpha)$ and quadratic local nature 
of maximum at $\alpha_0$ are also present. 
Characterizations such as ``exact" and~``proof" in this context are perhaps 
too strong and~premature.  

\vspace*{0.02in}
\noindent
{\em (iii)} Arguments of~\cite{Burmistrov:2022}, leading to claim of systematic 
effect in the analysis of~\cite{Horvath:2021zjk}, use numerical results 
of~\cite{Ujfalusi:2015a}. But systematics involved in the latter work is not 
discussed although smaller systems and smaller statistics were used. 
(Systematics in~\cite{Horvath:2021zjk} was also studied in 
Ref.~\cite{Horvath:2022ewv} albeit from a different angle.) 
Are there logs affecting standard MF analyses, appearing e.g. via
arbitrary replacement 
$p \!\equiv\! L^{-\alpha} \rightarrow p \!\equiv\! \text{const}\, L^{-\alpha}$.
Are they accounted for?

\vspace*{0.02in}
\noindent
{\em (iv)} The presence of $\log(L)$ powers in dimension estimates is always 
possible since they leave it intact. MF arguments are not necessary 
to invoke the possibility. But in the absence of a firm prediction, establishing 
their presence is a delicate numerical issue. Indeed, it is difficult to show that 
the influence of an unknown $\log$ is larger than that of subleading powers. 
Statistical strength of available data is frequently not sufficient to do that. 
Occam's razor approach is then an accepted resolution.

\vspace*{0.02in}
\noindent
{\em (v)} We strongly disagree with the comment {\em ``$d_\fir$ is nothing but f(d)"}.
There was also a calculus student who once declared 
{\em ``$\pi$ is nothing but $\,22/7 \!-\! \int_{0}^1 dx \,x^4(1-x)^4/(1+x^2)$"}. 
Perhaps there is a bit more to $d_\fir$ than the currently hypothetical 
$d_\fir \!=\! f_m(3)$. Its meaning is expressed in {\em (m1)}, being worked out 
in requisite detail by 
Refs.~\cite{Horvath:2018aap, Alexandru:2021pap, Horvath:2022ewv}. 

\vspace*{0.02in}
\noindent
{\em (vi)} Regarding comments on prospects for superuniversality in various 
dimensions, we repeat that our claim is only for $d_\fir$ at level better than 
two parts per mill in $D\!=\!3$. The gist of our message is that properties 
of space in Anderson transitions may be special relative to other 
characteristics. Scenario where superuniversality in $d_\fir$ is violated in 
$D\!=\!2+\epsilon$ but very accurate or exact in $D\!=\!3$ and higher 
dimensions is not contradictory in our opinion.

\medskip
\noindent
RESPONSE TO ADDENDUM

\smallskip
\noindent
New material was added to Ref.~\cite{Burmistrov:2022} in version 2. 
Its~focus is to suggest that inconsistencies found in originally proposed 
approximation \eqref{eq:805} of $\langle \efNm \rangle_L$ are rectified 
by the use of full integral formula~\cite{Burmistrov:2022}  
($f_m$ from moment method)
\begin{equation}
     \langle \efNm \rangle_L^m  \equiv 
     \sqrt{2 \ln(L) | f''_m(\alpha_0) | / \pi} \, \int_{-\infty}^D d\alpha \,L^{f_m(\alpha)}
     \label{eq:815}     
\end{equation}
More precisely, although it is admitted that $\langle \efNm \rangle_L^m$ is not 
a good theory of $\langle \efNm \rangle_L$, even in its full form \eqref{eq:815}, 
it is claimed to provide an accurate theory 
of $d_\fir(L,2)$~\cite{Horvath:2021zjk} 
(see~\eqref{eq:825} below). We will show that this is not the case when
associating $d_\fir$ with $f_m(3) \!=\! 2.733(3)$ as hypothesized
in~\cite{Burmistrov:2022}. 
In the second part we give further support to conclusions of 
Ref.~\cite{Horvath:2021zjk} 
and propose the resolution of inconsistencies between $d_\fir$ and
its representation via $f_m$, revealed by our analysis.


Discussion in ADDENDUM avoids elaborating on ramifications of standard 
MF variable choice $p(x) \equiv L^{-\alpha(x)}$, raised by our point {\em (ii)}. 
To be able to discuss related issues further, we thus adopt it as a 
{\em defining attribute} of MF approach in what follows. The framework
then seeks to determine the dimension function $f(\alpha)$ associated 
with $\alpha$-based partition of (lattice) space $\Omega(L)$. 
More precisely, let $A(\alpha,L,\psi)$ be the subset of points $x$ for which 
$\psi^+\psi(x) \!=\! L^{-\alpha}$ so that 
$\Omega(L) = \union_\alpha A(\alpha,L,\psi)$, and $\cP(\alpha,L,\psi)$
the count of these points~\cite{Burmistrov:2022}. 
Here $\psi$ is a critical Anderson state. Like in the UV case of fixed sets, 
$f(\alpha)$ aims to represent the dimension of set $A(\alpha)$. 
The stochastic nature of $A(\alpha,L,\psi)$ then forces the definition 
be based on IR scaling of average counts
$\cP(\alpha,L) \equiv \langle \cP(\alpha,L,\psi) \rangle$. We emphasize
that we now strictly distinguish $f(\alpha)$, constructed as above, from 
$f_m(\alpha)$ of the moment method which is its (sometimes exact) 
approximation. 

Given this setup, 
$\langle \efNm \rangle_L \!=\! \int d \alpha \, \cP(\alpha,L) \min\{1,L^{D-\alpha}\}$ 
of Ref.~\cite{Burmistrov:2022} holds, and $\langle \efNm \rangle_L$ is encoded 
in $\cP(\alpha,L)$. However, it is not fully encoded in $f(\alpha)$.
Similarly, the finite-$L$ IR dimension $d_\fir(L) \!\equiv\! d_\fir(L,s\!=\!2)$, 
namely~\cite{Horvath:2021zjk}
\begin{equation}
     d_\fir(L) \!=\! \frac{1}{\ln(2)} 
     \ln \frac{\langle \efNm \rangle_L}{\langle \efNm \rangle_{L/2}}
     \quad , \quad
     d_\fir \!=\! \lim_{L \to \infty} d_\fir(L) 
     \label{eq:825}     
\end{equation}  
is encoded in $\cP(\alpha,L)$ but not in $f(\alpha)$ alone. 
Before expressing the relationship of $f(\alpha)$ and $d_\fir$, we first 
emphasize that the definition of $d_\fir$ indicated in \eqref{eq:825} does not 
only apply to pure powers as claimed in~\cite{Burmistrov:2022}, but generally. 
Indeed, it returns $d_\fir$ for all
$\langle \efNm \rangle_L \!=\! L^{d_\fir}h(L)$ where $h(L)$ varies slower 
than any non-zero power near $L \!=\! \infty$, which is the usual meaning of 
Minkowski-like dimension. For example, linearly extrapolated $d_\fir(L)$ 
of~\cite{Horvath:2021zjk} corresponds to asymptotic behavior 
$L^{d_\fir} \exp(c/L)$.

Applying now the same rational to $f(\alpha)$ and its associated
$f(\alpha,L) \equiv f(\alpha,L,s\!=\!2)$ results in
\begin{equation}
    \langle \efNm \rangle_L \!=\! 
    \int_{-\infty}^{\infty} d \alpha \, v(\alpha,L) \,L^{f(\alpha,L)} \,\min\{1,L^{D-\alpha}\}      
    \label{eq:845}               
\end{equation}
Here functions $v(\alpha,L)$, $f(\alpha,L)$ are constructed from
\begin{equation}
    f(\alpha,L,\epsilon) = \frac{1}{\ln(2)}
    \,\ln \frac{\cP(\alpha,L,\epsilon)}{\cP(\alpha,L/2,\epsilon)}
    \label{eq:865}                   
\end{equation}
where $\cP(\alpha,L,\epsilon)$ is the count in interval $(\alpha-\epsilon,\alpha]$, 
via
\begin{equation}
    f(\alpha,L) \!=\! \lim_{\epsilon \to 0} f(\alpha,L,\epsilon) 
    \;\; , \;\; v(\alpha,L) \!=\! \lim_{\epsilon \to 0}
    \frac{\cP(\alpha,L,\epsilon)}{\epsilon\, L^{f(\alpha,L,\epsilon)}} \;  
    \label{eq:885}                       
\end{equation}
The rewrite \eqref{eq:845} of $\langle \efNm \rangle_L$ caters to 
MF variable $\alpha$ and is exact if limits in \eqref{eq:885} exist. 
In general, there may be non-analyticities already at finite $L$. 
The multifractal singularity spectrum is defined by
$f(\alpha) = \lim_{L \to \infty} f(\alpha,L)$. We emphasize again 
that $v(\alpha,L)$ that is finite and nonzero for all sufficiently large 
$L$ may still tend to $0$ or $\infty$ for $L \to \infty$, albeit slower than 
any non-zero power of $L$. 


Formula \eqref{eq:845} connects $d_\fir$ and $f(\alpha)$ most
generally. Under rather mild conditions on uniform convergence  
of $f(\alpha,L)$ it translates into
\begin{equation}
   d_\fir \!=  \sup\{\, \{f(\alpha) \!\mid\! \alpha \le D\}  \,\union\,
                  \{ f(\alpha) \!+\! D \!-\! \alpha \!\mid\! \alpha > D\} \,\} \;
    \label{eq:905}                                     
\end{equation}
This representation applies in any context with well-defined $f(\alpha)$
and it is a nice contribution of Ref.~\cite{Burmistrov:2022} to spark this 
connection. However, using 
it in practice amounts to ``putting the cart before the horse" unless 
a precise $f(\alpha)$ is available, which is very rare.


For 3D Anderson criticality problem, it is reasonable to expect that 
\eqref{eq:905} will eventually yield $d_\fir = f(3)$, but $f(\alpha)$ has not 
been computed yet. Rather, the MF framework is replaced by an approximation 
afforded by the moment method. Such mMF formalism substitutes $f(\alpha)$ 
with $f_m(\alpha)$ defined by generalized dimensions via Legendre transform, 
and assigns identical weight to distinct $\alpha$-populations, namely
\begin{equation}
      v(\alpha,L)\, L^{f(\alpha,L)}  \quad \longrightarrow\quad  
      \cC(L)\, L^{f_m(\alpha)}
      \label{eq:925}                                            
\end{equation}
in all expressions, including Eq.~\eqref{eq:845}. For this to be consistent, 
some features that are automatic in MF have to be explicitly imposed in 
mMF, in particular the normalization of counts and the wave function
\begin{equation}
      \frac{1}{\cC(L)} = 
      \int_{-\infty}^\infty d\alpha\, L^{f_m(\alpha) - D} = 
      \int_{-\infty}^\infty d\alpha\, L^{f_m(\alpha) - \alpha}     
      \label{eq:945}                                                   
\end{equation}
Arguments in~\cite{Burmistrov:2022} leading to $d_\fir \!=\! f_m(D)$ also rely 
on~\cite{Mirlin_2006} 
\begin{equation}
     f_m(2D-\alpha) = f_m(\alpha) + D - \alpha
     \label{eq:985}                                                                       
\end{equation}
which thus has to be imposed in this particular analysis as well. We can now 
discuss the merits of mMF approximations to $d_\fir(L)$.

\begin{figure}[t]
   \includegraphics[width=0.42\textwidth]{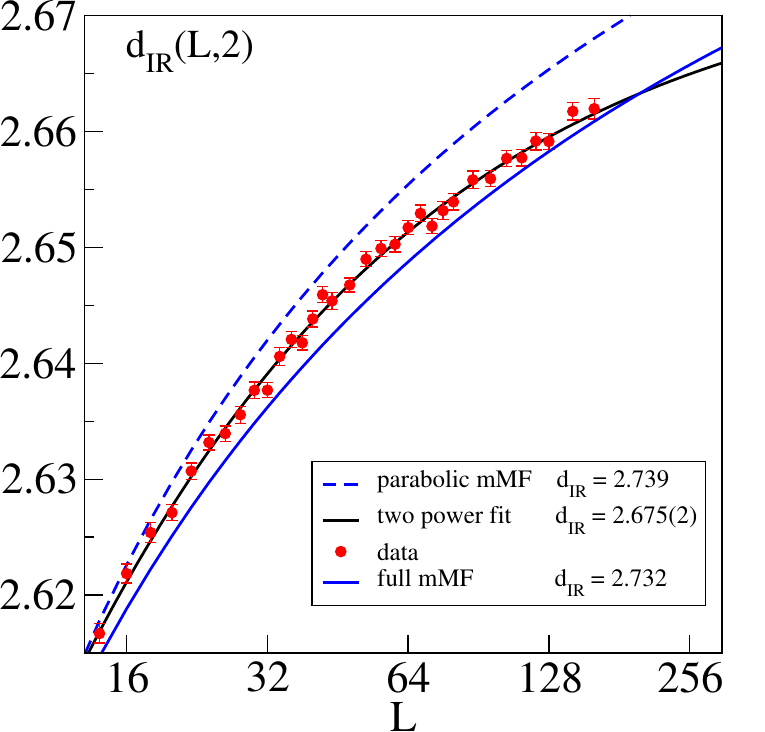}
   \vskip -0.05in   
   \caption{Parabolic (dashed blue) and full (blue) mMF predictions for $d_\fir(L)$ 
   vs true values (data). Dependence from 2-power fits of  $\langle \efNm \rangle_L$ 
   in corresponding range of $L$ is also shown.}  
   \label{fig:Burm2}
   \vskip -0.16in
\end{figure}

\noindent
{\bf Parabolic mMF.} It is easy to check that, given $\alpha_0$, the only
parabolic $f_m(\alpha)$ satisfying \eqref{eq:985} is
\begin{equation}
     f_m^p(\alpha) =  D - \frac{1}{4}\frac{(\alpha - \alpha_0)^2}{\alpha_0-D}
     \quad,\quad \alpha_0 > D
     \label{eq:1005}                                                                            
\end{equation}
which also satisfies conditions \eqref{eq:945} at each $L$. Therefore
\begin{equation}
     f_m^p(D) = D - \frac{\alpha_0-D}{4}   \;\longrightarrow\;  f_m^p(3)=2.739
     \label{eq:1045}                                                                                 
\end{equation}
for $D\!=\!3$ and $\alpha_0\!=\!4.043$. This is thus the prediction for $d_\fir$ by
mMF in parabolic approximation. Note that $\alpha_0$ is quoted 
in~\cite{Burmistrov:2022} without error. The associated $d_\fir(L)$ given by 
master mMF formula \eqref{eq:815} is shown in Fig.\ref{fig:Burm2} together with 
numerical data. We have extended the range of studied lattices and their 
statistics for this purpose. The obvious disagreement of the theory with the data 
makes the parabolic mMF prediction $d_\fir \!=\! 2.739$ unreliable. 


\smallskip\noindent
{\bf Full mMF.}  The full mMF procedure first parametrizes $f_m(\alpha)$ 
using numerical data from the moment method, and then uses formula 
\eqref{eq:815} to predict $d_\fir(L)$. Eqs.~\eqref{eq:945}, \eqref{eq:985} 
have to be satisfied with sufficient accuracy so that consistency of 
the prediction is not compromised. 

It is important to point out that, unlike in the case of parabolic mMF,
$f_m(3)$ is not predicted here but rather introduced by hand. Indeed, 
parametrization of mMF data obviously reproduces mMF values. 
Thus, in terms of verifying the hypothesis $d_\fir \!=\! f_m(3) \!=\! 2.733(3)$,
the procedure is a tautology. Its actual meaning is to verify the consistency 
between the MF approach, which is exact even at finite $L$ (formulas 
\eqref{eq:845} and \eqref{eq:905}), and the mMF approach which only 
provides an approximation. 

To obtain the full mMF prediction for $d_\fir(L)$, Ref.~\cite{Burmistrov:2022} 
uses the data of Ref.~\cite{Ujfalusi:2015a} to parametrize $f_m(\alpha)$ as
\begin{equation}
    3 - 0.2662(\alpha \!-\! \alpha_0)^2 \!- 
    0.0254(\alpha \!-\!\alpha_0)^3 \!- 0.0061(\alpha \!-\!\alpha_0)^4  
     \label{eq:1085}                                                                                     
\end{equation}
The errors of parameters involved were not given. The confrontation 
with data, which is equivalent to comparing mMF prediction for $d_\fir(L)$ 
to MF prediction, is shown in Fig.\ref{fig:Burm2}. Large disagreement 
is seen not only at computed $L$, but also in the observed trends. We are 
thus led to conclude that, like its parabolic approximation, even the full mMF 
prediction $d_\fir \!=\! f_m(3) \!=\! 2.733(3)$ is unreliable.  


Before proceeding to provide other evidence favoring conclusions of 
Ref.~\cite{Horvath:2021zjk}, and to analyze the origin of observed 
inconsistencies, we point out that one cannot use the simplified 
quadratic form (Eq.~(12) in~\cite{Burmistrov:2022}), to fit for $f_m(3)$, 
$f_m''(3)$ and $f_m''(\alpha_0)$. These values are fixed by $f_m(\alpha)$ 
and releasing them leads to gross violation of consistency conditions 
\eqref{eq:945} and \eqref{eq:985}. Dropping cubic and quartic terms 
in~\eqref{eq:1085} already violates them to a dangerous degree, as seen 
by comparing to consistent parabolic approximation~\eqref{eq:1005}. 
The same applies to additional rescaling of $L$ speculated upon in 
the ADDENDUM.  

\begin{figure}[t]
   \includegraphics[width=0.42\textwidth]{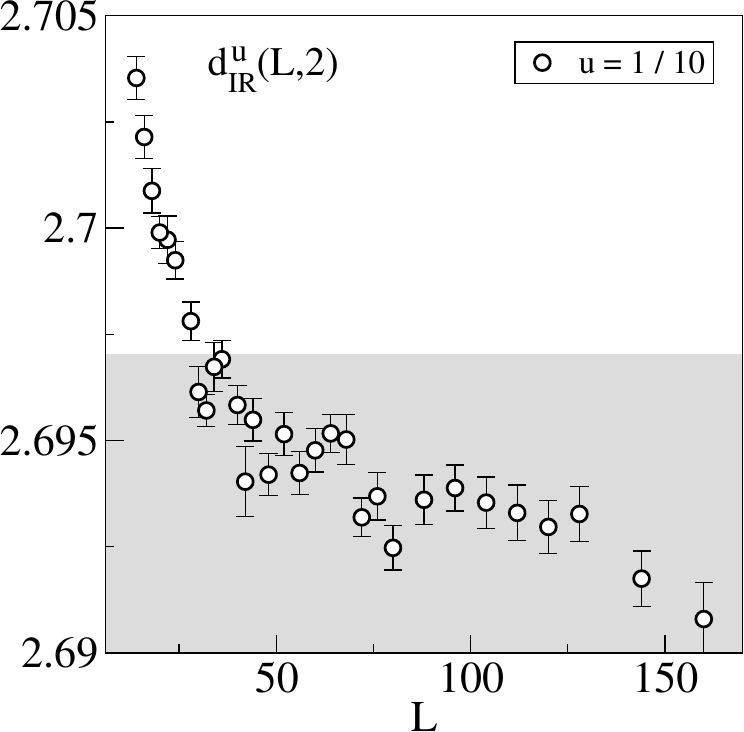}
   \vskip -0.05in   
   \caption{Function $d_\fir^u(L)$ is decreasing for $u=1/10$, constraining  
   $d_\fir$ to be below the onset of shaded region.}  
   \label{fig:Burm3}
   \vskip -0.16in
\end{figure}

\medskip
\noindent
PROPOSED RESOLUTION
\medskip
\noindent

\noindent
Below we present additional features of our data which, together with 
the above, point to coherent explanation of observed inconsistencies 
between the mMF prediction and numerical results for $d_\fir$.

\noindent
{\bf 1. Upper Constraint.} Construction of effective counting dimension
($d_\fir$ here)~\cite{Alexandru:2021pap, Horvath:2022ewv} from effective 
number theory~\cite{Horvath:2018aap} includes identifying counting 
schemes that lead to consistent effective supports. These schemes, 
represented by functions of $P=(p_1,\ldots,p_\nrN)$, are specified
 by~\cite{Horvath:2022ewv} 
\begin{equation}
   \efN_{(u)}[P] = \sum_{i=1}^\nrN 
   \min\, \Bigl\{ \frac{ \nrN p_i}{u}, 1 \Bigr\}    \quad,\quad
   0 < u \le 1   
   \; \label{eq:1135}         
\end{equation}
Note that $\efN_{(1)} \!=\!\efNm$. Uniqueness of $d_\fir$, proved 
in~\cite{Horvath:2022ewv}, means that the associated $d_\fir^u(L)$
satisfy
\begin{equation}
    \lim_{L \to \infty} d_\fir^u(L) = d_\fir  \quad,\quad \forall\; u 
    \; \label{eq:1155}              
\end{equation}  
This property can be used to constrain $d_\fir$ based on robust 
behavior of data rather than fitting. Indeed, from
$\efN_{(u_1)}[P] \ge \efN_{(u_2)}[P]$ for all $u_1 \!<\! u_2$ and $P$,  
it follows that $d_\fir^{u_1}(L) \ge d_\fir^{u_2}(L)$ for sufficiently large 
$L$. This implied order was already seen even on very small Anderson
systems~\cite{Horvath:2022ewv}. Hence, although 
$d_\fir(L)$ approaches $d_\fir$ from below, $d_\fir^u(L)$ may descend 
from above for sufficiently small $u$, and majorize the value of the limit. 
In Fig.~\ref{fig:Burm3} we show this for $u\!=\!1/10$ which is in 
a decreasing~regime. The observed trend suggests a (conservative) 
upper~restriction
\begin{equation} 
     d_\fir < 2.697  \ll 2.733(3)
     \; \label{eq:1175}                       
\end{equation}   
where ``$\ll$" refers to differences relevant in this problem. 

\noindent   
{\bf 2. The Log.}  At the heart of arguments 
in~\cite{Burmistrov:2022}  is the claim that the leading power governing 
$\langle \efNm \rangle_L$ has a logarithmic prefactor, i.e.  
$\langle \efNm \rangle_L \propto \ln(L)^\gamma L^{d_\fir}$ for $L \!\to\! \infty$. 
While the strict mMF prediction $\gamma\!=\!-1/2$ is not visible in the data
(see our first response), it is prudent to inquire about the possibility of other 
non-zero value since it is allowed on general grounds. To that effect, we fitted 
our data over the entire available range ($6 \le L \le 160$), and obtained a very 
good description ($\chi^2/\text{dof}\!=\!1.0$) with parameters
\begin{equation} 
  \text{log:} \quad\;\;  d_\fir = 2.704(1) \quad\; , \quad\;
      \gamma =  -0.202(2) 
      \;\; \label{eq:1195}                       
\end{equation}   
While this nominally suggests that the obtained $\gamma$ and $d_\fir$ 
characterize true asymptotics, the tension of the latter with 
constraint~\eqref{eq:1175} and Fig.~\ref{fig:Burm3} makes it somewhat 
dubious. Also, logs are known for their capacity to subsume other behaviors. 
Needless to say, description \eqref{eq:1195} also contradicts mMF 
predictions. 

\noindent   
{\bf 3. The Powers.} 
However, barring standard mMF expectations, there is no fundamental 
reason for multiplicative log in the leading term. In fact, we obtained 
a high-quality fit ($\chi^2/\text{dof}\!=\!1.0$) of our extended 
$\efNm$ data over the full range ($6 \le L \le 160$) for a 2-power 
description $aL^{d_\fir} + bL^{d_m}$ with  
\begin{equation} 
   \text{powers:} \quad\;\;  
   d_\fir = 2.673(2) \quad\; , \quad\;  d_m =  1.998(29) 
   \;\; \label{eq:1215}                       
\end{equation} 
Note that there is no tension of this log-free description with 
constraint~\eqref{eq:1175} and Fig.~\ref{fig:Burm3}. While this 
already provides rationale for adopting scenario~\eqref{eq:1215} 
over~\eqref{eq:1195}, the logs are deeply rooted in mMF practice 
and their absence needs an explanation in multifractal terms.  

\noindent   
{\bf 4. The Resolution.}  Described inconsistencies between mMF 
predictions and data have two principal manifestations, both pointing 
to the same underlying cause. First, they suggest in distinct ways 
that $d_\fir \!\ll\! f_m(3) \!=\!2.733(3)$. At the same time, relation 
\eqref{eq:905} clearly shows that $f(\alpha)$ is connected to $d_\fir$, 
likely via $d_\fir=f(3)$ i.e. in the way proposed in~\cite{Burmistrov:2022} 
but with $f_m$ replaced by $f$. This suggests that the ensuing tension 
is connected to possible differences between these characteristics.

Secondly, our numerical analysis raised the possibility that multiplicative
log may not accompany the leading power in $\langle \efNm \rangle_L$. 
At the same time, the ``log paradigm" in mMF stems in fact from its 
``shape paradigm". Indeed, $f_m(\alpha)$ is expected to be a smooth 
concave function with special points, most notably a single $\alpha_1$ 
satisfying $f_m(\alpha_1) \!=\! \alpha_1$, and a unique $\alpha_0$ 
satisfying $f_m(\alpha_0)\!=\!3$ and $f_m'(\alpha_0)\!=\!0$. In this 
setting, the relevant $\alpha$-integrals are dominated by one spectral point 
and multiplicative logs in leading powers are almost inevitable indeed.  
This again steers us in the direction that $f(\alpha)$, which does not
have such restrictions, may be different from $f_m(\alpha)$ in a manner 
that avoids the leading multiplicative log. 

\begin{figure}[t]
   \includegraphics[width=0.42\textwidth]{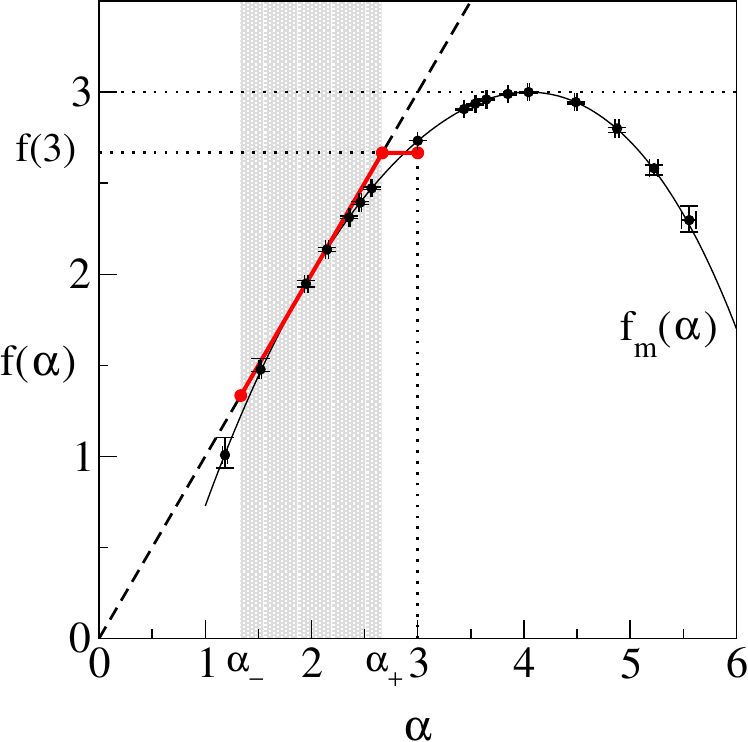}
   \vskip -0.05in   
   \caption{Possible realization of $f(\alpha) \ne f_m(\alpha)$ is via $f(\alpha)$
   featuring the red solid segment. The $f_m(\alpha)$, shown to guide the eye, 
   is from Ref.~\cite{Ujfalusi:2015a}.}  
   \label{fig:Burm4}
   \vskip -0.16in
\end{figure}

Following the above leads, we propose that at Anderson criticality in 3D,
functions $f_m(\alpha)$ and $f(\alpha)$ are not identically equal to each
other, namely
\begin{equation}
   \text{conjecture:} \quad f  \neq f_m   \quad , \quad\
    d_\fir = f(3) < f_m(3)   \;\;
    \label{eq:1235}                           
\end{equation}
One possible realization commensurate with existing evidence is 
the presence of a segment in $f(\alpha)$ represented by the red 
solid line in Fig.~\ref{fig:Burm4}. In particular, $f(\alpha)=f(3)$ for 
$\alpha_+ \le \alpha \le 3$ and $f(\alpha)=\alpha$ for 
$\alpha_- \le \alpha \le \alpha_+$. The horizontal segment is key 
for the present argument because it can generate pure power 
$aL^{f(3)} = aL^{d_\fir}$ upon evaluating its contribution to 
\eqref{eq:845}. Indeed, the integral determining $a$, namely 
$\int_{\alpha_+}^3 v(\alpha,L) d \alpha$, cannot diverge for 
$L\to \infty$ and our numerical experiments show that it is 
increasing and thus finite. While details will be given elsewhere,
the horizontal segment offers a robust mechanism for generating
a log-free leading power in $f(\alpha)$. Note that the difference that 
sparked this discussion is one between the red dot and the black 
data point at $\alpha=3$ in Fig.~\ref{fig:Burm4}.

The nominal purpose of the linear $(\alpha_-, \alpha_+)$ segment is to ensure
that there is a gap between $f(3)=d_\fir$ and the subleading behavior,
which is needed to explain the remarkably accurate 2-power description
\eqref{eq:1215} of the data. Indeed, one can easily check that, for slowly
varying (in $\alpha$) $v(\alpha,L)$, the integral \eqref{eq:845} only generates
powers of $f(3)=\alpha_+$ and $\alpha_-$, with prefactors vanishing as 
$1/\sqrt{L}$. This is both harmless to the leading term and creates~a~gap.

The attractive feature of the above scenario is that it is also consistent
with the recent multidimensional analysis~\cite{Horvath:2022klk} which   
suggests that there is a finite probability to find dimensions in any part of 
interval $[d_\smin,d_\smax] \approx [4/3,8/3]$. This contradicts the usual 
mMF paradigm that, in the thermodynamic limit, only information dimension 
$f_m(\alpha_1)$ is found. The corresponding difference is exactly the one 
between $f_m(\alpha)$ only having a single point satisfying 
$f_m(\alpha) \!=\!\alpha$, and the above proposal that $f(\alpha)\!=\!\alpha$ 
on the entire interval $[\alpha_-,\alpha_+]$. In fact, the connection is direct
and $[d_\smin,d_\smax] \equiv [\alpha_-,\alpha_+]$. 

Another surprising support for the above proposal comes from the details 
of our findings in Ref.~\cite{Horvath:2022klk}. Indeed, given that 
$\alpha_-\!=\!d_\smin \!\approx\! 4/3$, it may seem surprising that the 2-power 
description \eqref{eq:1215} involves a subleading $d_m \!\approx\! 2$. 
However, this is in fact expected given the results of multidimensional 
analysis~\cite{Horvath:2022klk}, which concluded that dimension 
$d_m \!\approx\! 2$ is discrete. Its special status is not captured by 
$f(\alpha)$ since it has a $\delta$-function prefactor. But its role is of course 
crucial for capturing the small $L$ behavior which is exactly why 2-power 
description works so well. In fact, the closeness of fitted value to $d_m \!=\!2$ 
independently supports the proposition of~\cite{Horvath:2022klk} that 
$d_m$ may be an integer.

\smallskip
\noindent
We finally wish to emphasize three points.
\smallskip

\noindent {\bf (a)} 
By performing an extensive numerical analysis, we have shown that, 
even in the full form \eqref{eq:815} of Ref.~\cite{Burmistrov:2022}, mMF 
is unable to describe data for $d_\fir(L)$. Observing the practice of not 
blaming accurate data for refusing to follow the theory but the other way 
around, we are led to conclude that the mMF prediction 
$d_\fir \!=\! f_m(3) \!=\! 2.733(3)$ is unreliable. As shown from several different 
angles, the extent of the tension is such that the conclusions on $d_\fir$ based 
on mMF, at least in the form suggested by~\cite{Burmistrov:2022}, cannot be 
used to judge the correctness of our conclusions {\em (m1)} and 
{\em (m2)}~\cite{Horvath:2021zjk}.

\smallskip

\noindent {\bf (b)} 
Our analysis showed that the observed discrepancies have to be interpreted,
in fact, as those between predictions of MF and mMF. This observation is 
crucial for the proposed resolution $f \!\neq\! f_m$ \eqref{eq:1235}. In that 
regard, the contribution of Comment~\cite{Burmistrov:2022} is very valuable. 
While the debate regarding $d_\fir$ and $f_m(3)$ may continue, 
it already focused the attention on the possibility to evaluate $f(\alpha)$ directly. 
This may not be unrealistic since the computer power has improved greatly 
from early days of MF when similar calculations were attempted. Nevertheless, 
computing $f(\alpha)$ is certainly not an efficient way to evaluate a basic 
geometric characteristic of a multifractal such as~$d_\fir$. 

\smallskip

\noindent {\bf (c)} Our proposal that $f(\alpha)$ features a finite 
$f(\alpha) \!=\!\alpha$ segment is equivalent to the conclusion of 
Ref.~\cite{Horvath:2022klk} that Anderson critical states are not only 
multifractal, but also multidimensional. This possibility entails an important
geometric consequence, namely that critical Anderson states may not
be self-similar. Indeed, in standard UV settings, self-similar multifractals 
have to obey the shape paradigm~\cite{falconer2014fractal}. 
If multidimensionality is indeed realized, this would bring new and interesting 
geometric detail into the long story of Anderson transitions, and MF approach 
may play a very relevant corroborating role. 

\medskip

\begin{acknowledgments}
   P.M. was supported by Slovak Grant Agency VEGA, Project n. 1/0101/20.
   I.H. is indebted to R.~Mendris for many discussions.
\end{acknowledgments}

\bibliography{my-references}

\end{document}